# Power Systems of the Future

Mario Rabinowitz
Armor Research, 715 Lakemead Way, Redwood City, CA 94062-3922
Mario715@earthlink.net

**Abstract**

Electric power is a vital ingredient of modern society. This paper in conjunction with previous papers was written to provide an insight into the physics and engineering that go into electric power systems and their modernization. Topics covered here are Direct Current; Superconducting Generators**;** Energy Storage; Voltage Sags; Grid Stability, Power System Planning and Operations; Biological Effects of Electromagnetic Fields; Dispersed Generation; Information Superhighway Synergy; Distribution Automation; Conclusion.

## Direct Current, Grid Stability, and Superconducting Generators

For long distance transmission, dc is significantly less costly than ac. Direct current also serves an important function in tying together two large ac systems even for short distances, thus circumventing problems related to synchronization. However for ordinary transmission, dc is limited to long lengths because of the high cost of rectification and inversion at the two ends of the line involving high power diodes, transformers, and filters. Substantial filtering is typically required, at significant expense, to remove the ripple at the dc output. Substantial filtering is also required on the primary side of the transformer to prevent surge and harmonics from getting back to the generator.

An increase in frequency reduces the degree of ripple and of filtering required. Mere substitution of a conventional low-voltage generator that provides high frequency is unsatisfactory, since this would increase the transformer impedance and losses (especially core losses), and decrease the power transmitted. Furthermore, it is impractical to try to generate at substantially higher voltages with a conventional generator. This is because the flux density is limited by iron saturation and the armature turns must be insulated from the grounded iron, thereby limiting the ampere-turns density and the voltage.

The amplitude of the ripple is decreased approximately inversely as the square of the number of phases. Similarly, the filtering requirement is reduced as the frequency is increased. In a conventional system utilizing a high-voltage transformer, as the number of phases are increased, the cost increases proportionately, which is substantial. Development of a high-voltage superconducting generator will entirely eliminate the transformer. This permits both an increase in frequency and in the number of phases without a concurrent cost penalty. This means that less reactance will be needed in both the input and output sides of the diodes. An increase in the frequency by a factor of 3 can reduce the reactor requirement by a factor of 3. An increase in the number of phases by a factor of 2 reduces the reactor requirement by a factor of 4.

The per-unit synchronous reactance of a superconducting generator is about 1/4 that of a conventional generator of similar rating. This results in an increased steady-state stability limit of the superconducting machine by as much as a factor of four when the transmission line reactance is relatively small. Superconducting power generation and superconducting magnetic energy storage (SMES) were covered in-depth in the May 2000 issue of Power Engineering Review, so only a little additional material is covered in this article.

**Energy Storage, Voltage Sags, and Grid Stability**

The increasing scarcity of the earth's primary energy resources has steadily raised their cost, and in turn the cost of electricity production. This has led to increased interest in load-leveling energy storage systems. The procedure would be to store excess energy that is generated during off-peak periods when the load is low, and later to deliver it during peak load periods. There are a number of energy storage methodologies such as pumped-hydro, compressed-air, flywheel, thermal, magnetic, and electro-chemical storage such as batteries and fuel cells. The interest in SMES is not only because SMES is expected to have a higher round-trip efficiency than any other large-scale storage technology, but because it can also be used to help stabilize the power grid by quickly adding or removing power.

The use-capacity of an ac line may be increased by rapid regulation of a SMES unit [1]. The ac line use-capacity primarily depends on three factors: 1) transient stability limit, 2) voltage collapse, 3) electro-mechanical oscillations resulting from subsynchronous resonance of the turbine shaft and the transmission network. Of these, SMES is expected to most easily ameliorate 3) by damping the oscillations. Smaller SMES coils are contemplated where their primary use would be in maintaining power system stability [2].

A small coil of 8 kWh (30 MJ) storage capacity and 10 MW peak power was successfully tested on the Bonneville Power System for 6 months in 1984.[3] It was used to damp subsynchronous (< 60 Hz) oscillations on the Pacific Coast AC/DC transmission line that connects Washington state to southern California. Although much development needs to be done to make SMES economically competitive with other energy storage schemes, it has no peer for the combined functions of energy storage and grid stability.

There is even a POWER QUALITY niche for what may be called microSMES (MSMES), despite the fact that it is very inefficient and very costly. The commercial viability of this technology is readily apparent in the semiconductor industry where voltage sags of 2 or 3 cycles can ruin several single silicon crystal ingots, each worth $15,000. Such power dips costing between $50,000 and $100,00 implies that the cost of a MSMES can be recovered in about 3 years. However, they would prefer to pay the utility a lot more per kW-hr than normal cost, because in the silicon wafer fabrication industry, equipment and buildings become obsolete in less than 10 years. Advocates estimate that MSMES could save U.S. industry $12 billion per year.

More research needs to be done for better HTSC in wire form before they are adequate for SMES. However, it may be possible to circumvent this problem by using HTSC in a wide variety of aggregate forms such as granules, particulates, foil, and thin film in which the magnetic energy is stored in trapped form to be released as electrical energy by magnetically coupling to a normal coil as the trapped field is caused to decay [4, 5]. This trapped-field energy storage (TES) has the advantages of elevated temperature operation, that a HTSC wire coil need not be made, and the elimination of lossy leads. However, much research needs to be done to make TES a practical reality.

**Power System Planning and Operations**

In the future, electric utilities will find themselves in an increasingly competitive environment which will push the power system to the limits of its operability. Conflicting requirements of operating the system closer to its thermal and stability limits, responding quickly to wholesale energy transactions, and yet maintaining system security and integrity will demand well-coordinated power system planning, dispatching, and operations. As the quantity of wheeled power increases together with the rate of wheeling transactions, the system vulnerability will also increase. There is much ongoing work to develop methods, procedures, guidelines, and software products to deal with these contingencies in the operation and control of the grid.

If these programs are successful in achieving at least a modest 5% savings in power production, the overall nationwide impact is significant. With a roughly $50 billion/year fuel cost, the potential savings amounts to $2.5 billion/year. Goals of this work are to enable planners and dispatchers of the future to:

- Determine safe transfer limits across critical interfaces so as not to exceed stability and thermal limits.

- Determine how much margin is available on generation and delivery systems.
- Determine how best to control the delivery system.
- Determine how best to respond both judiciously and expeditiously to constantly changing transaction opportunities.
- Determine how best to predict, implement, and supervene over transaction decisions.

**Biological Effects of Electromagnetic Fields**

The issue of adverse biological effects of electromagnetic fields (EMF) will crucially affect power delivery of the future. At this point it is not clear whether or not there is strong evidence for adverse effects. There are clearly some beneficial effects, and a number of effects that are neither beneficial nor adverse. Electric utilities cannot afford to be complacent on this issue.

Biological systems respond to unbelievably low electric and magnetic fields. Freshwater catfish respond to electric fields as low as $10^{-6}$ V/cm. Marine sharks and rays are sensitive to less than $5 \times 10^{-9}$ V/cm, and use this sensitivity to navigate using the voltage gradients induced by ocean currents flowing in the earth's magnetic field. Magnetite was discovered in the human brain in 1992. Strands of magnetite function like compass needles to help one-celled bacteria navigate. Magnetite has been found in homing pigeons, salmon, dolphins, tuna, bats, and honeybees, and may be part of their navigational systems.

The application of magnetic fields to broken bones has clearly been shown to speed up the mending of bones. B. F. Sisken et al have studied the basic problems encountered in nerve injury and regeneration [6]. They find that electric and electromagnetic fields may help the healing process. They have demonstrated that pulsed electromagnetic fields accelerate nerve regeneration in the injured sciatic nerve of rats, and that this has broad implications for the clinical use of these fields in the management of nerve injuries [7].

On the other hand, in 1991 Adair wrote a paper whose object is to show that normally encountered 60 Hz electromagnetic fields have no significant biological effect at the cell level [8]. More recently Bennett came to the same conclusion [9]. These are very well-written papers that make a strong case that the electric and magnetic fields from power lines are well below the levels of natural exposure. Aside from a few inconsequential errrors, their physics, calculations, and numbers appear reasonable.

However they consider only single cell response. They do not adequately address the point that an organized system of living cells is sensitive to and can respond to a much smaller signal than that of a single cell. This is analogous to digital processing in finding a signal in a situation where there is a large noise-to-signal ratio. While the work of Weaver and Astumian [10] is referenced, they do not seem to adequately respond to the point that living cells can and do react to field levels well below the thermal noise limit.

Biological molecules made of long chains of amino acids fold themselves up properly in an instant. Their dynamics can be as short as a femtosecond ($10^{-15}$sec). Bennett focuses only on the breakup of a biological molecule, while neglecting the effects of interference with the folding process. Thus he thinks that only much larger fields can be significant biologically.

Scientists at Sandia National Labs and the University of New Mexico claim to have observed unambiguous and reproducible non-thermal deleterious effects of pulsed magnetic fields on developing quail embryos. The exposed group had 10 times the abnormalities compared to a control group. They used larger fields than usually encountered to enable them to produce clear-cut, easily observable effects where the effect is not lost in the noise. Bennett and Adair did not cover these experiments.

Most recently, Blackman et al claim to have unequivocally and reproducibly demonstrated that nerve growth can be inhibited by exposure to power line levels of magnetic fields as a function of either magnetic flux density or frequency [11]. These tests were in vitro rather than in vivo. Though nerve repair has been verified in vivo, it remains to be shown whether nerve growth inhibition can be demonstrated in living animals. Reproduction of these experiments has not yet been reported in other laboratories. If reproducible, their results should be judged independently of their Ion Parametric Resonance Model, which is used to guide their research and to explain their results.

Straightforward ways can be implemented to minimize EMF concerns with respect to transmission lines, distribution lines, and substations. Increasing the number of phases decreases the electromagnetic field. Generally balanced 3-phase underground cable has a smaller EMF than balanced 3-phase overhead lines. This is simply because the solid insulation has higher dielectric strength than air, permitting closer spacing of the 3 phases and a faster fall-off of the fields from the lines. Unbalanced lines, and ground loops for single phase circuits produce large electromagnetic fields. Epidemiological studies indicate that if there is a causal connection between EMF and adverse biological effects, it is very tenuous. Thus one may be optimistic that we can find reasonable corrections if we do find causal connections.

**Dispersed Generation**

Traditionally, both in generation and in transmission, electric utilities have pursued economies of scale with large power plants in increasing efficiency and in reducing capital and operating costs. However, both natural and artificial constraints limit such expansion. As discussed earlier, transmission line capacity is limited by a practical line voltage of 1200 kV. Temperature and pressure limits are being approached in turbine and boiler design. Sometimes the reason for the limit is clear, but often the limit constraint is quite subtle.

When these limits are reached, new technological breakthroughs such as the superconducting generator can result in new greater limits, or to new ways of doing business. This may be as simple as a new way of generating electricity; or with even farther reaching consequences, utilities may have to alter the way they are structured, plan, and operate. Even if the physical limits were far greater than they are now, such high power levels would be beyond what a utility would want to risk on a single machine or power line. Although load growth in general may be forecast fairly accurately, due to re-regulation, load growth for individual utilities may be fraught with uncertainty. Therefore even at the present limits, with such uncertainty in load growth, building large capital intensive plants for the purpose of economy of scale is a very risky undertaking.

Dispersed generation in its manifold manifestations of co-generation, wheeling, renewable generation, fuel cells, etc. must be properly considered not only because of potential competition, but because they may afford new opportunities for utilities. Can future load growth be met by alternate strategies to new central plant construction or upgrade, and/or new transmission and distribution system erection or upgrade?

The trend of the electric power industry has been to increase utilities' relative investment in transmission and distribution (T & D). Electrical World [12] estimated that the T & D share of new utility investment increased to 80% in 1997. In part, this may be due to a present excess in generation capacity. However, it may also reflect the fact that many new technologies for generation, such as fuel and solar cells, may become commercial within the next few decades making it imprudent to invest in additional conventional generation. Windpowers' present commercial success is due in part to serendipitous pricing contracts made during the OPEC-created oil crisis of the '70s. As we shall next see, if the scale is not too modest, small gas-fired turbo-generators are already commercially viable.

Stanford University is a prime example of the kind of dispersed generation that may become more prevalent in the next few decades. In 1987 General Electric installed a 50 MVA gas-fired steam turbo-generator power plant on the Stanford campus to replace the power supplied by the local electric utility.  This plant generates the electricity and steam heat for all of Stanford's needs (including the Stanford hospital) by burning methane delivered by Pacific Gas and Electric.  Stanford sells its excess power to PG & E.  This facility has saved (earned) Stanford millions of dollars per year.

While 50 MVA is small compared to 1500 MVA, or even 300 MVA, it is probably big enough to gain some of the advantages of economy of scale.  Clearly 50 MVA is a bit above the threshold for commercial viability, at least for the California market, and probably for an even larger segment of the U.S. market.  We may be surprised to find that 50 MVA is economically viable across the entire United States, and that the threshold may be lower than 20 MVA in California.  Such facilities, preferably in cogen form, but even if they don't use waste steam, will very likely see increasing usage in the next 20 years.  Whether they are owned and operated by utilities, or by opportunistic entrepreneurs, is a business decision that utilities make based upon each individual utility's preferences, assets, and needs.

At this time all the non-combustible renewable resources of sun, earth, and moon power represent less than 1% (< 7000 MW) of the power generated in the U.S., rather than the up to 5% hoped for before 1980. Just because renewable energy sources like the sun (solar thermal, photovoltaic, windpower, ocean thermal gradients), earth (geothermal), and the moon (tides) did not prove to be commercially viable in the past, is no reason to think that this will always be the case.  The need for renewable resources becomes manifestly clear in realizing that even if the earth were a hollow sphere full of ready-to-use oil, it would be depleted in a few centuries -- given the present rates of use and of increase in use.  Of course, renewable is a relative term depending on time scale, as even the sun will eventually burn out.

So it is clear, that although dispersed generation using renewable sources may not make a large impact in the next two decades, it is certainly the way of the future. Shell International Petroleum predicts that renewable power will dominate world energy production by the year 2050.  This appears overly optimistic, but their prediction of an oil crisis in the 1970's appeared  overly pessimistic until it happened.  Oil and gas companies are not standing by idly, and may be expected to lower prices and find ways to burn these fuels more cleanly.  The inevitable can be put off, but not indefinitely.

Presently the costs of fuel and photovoltaic cells, solar thermal, and possibly windpower are too high to consider them seriously as contenders on the economic playing field.  However in the future, we may not be competing on a level playing field. These technologies are viewed as environmentally more acceptable than fuel-burning central-power plants.  Utilities should be prepared for further preference points to be given to these new developing technologies by public utility regulators faced with smog-filled cities like Los Angeles, Mexico City, New Delhi, and Beijing. When these technologies come into vogue, whether they are implemented by utilities or newcomers, we need to know potential impact on existing and future distribution, system protection, substation and transmission facilities. Generation determines how we do T&D.

**Information Superhighway (IS) Synergy**

Reliable and timely information is a valuable commodity. The networking of individuals, teams, associations, companies, and corporations has developed a need for more efficient exchange of information locally and globally. The main driver is speed of access, though it may be quite some time before much can be done about speed of assimilation. This requires technical innovations to be made in a broad spectrum of scientific disciplines including microwave transmission and reception, waveguides, optical fibers, and a synergy between optical fibers and power lines. Essentially, the information superhighway (IS) is a network of communication systems providing high speed, broad-band integrated services. Thus the new telecommunications and information technologies of the emerging IS present electric utilities with a new set of challenges.

There is great interest in combining power lines with fiber optic cables to also carry telecommunications as part of the future communications superhighway. It should not be taken for granted that fiber optic cables will be trouble free on high voltage overhead lines. Over a long period of time, the effects of electric stress and high voltage corona can degrade an unshielded fiber optics cable if the fibers are exposed to a high electric field. (This would not be a problem outside the ground sheath of an underground power cable, as there is no electric field there from the high voltage line.) Although the grounded shield wires of an overhead line have a much lower electric field environment than the power lines, lightning is more likely to strike the grounded shield lines and damage the optical fibers. It is not always possible to put optical fiber cables on low voltage phone lines. In the case of long distances, there are no phone lines on which the fiber optic cable can be carried because distant phone transmission is by microwaves.

Though it may not be expected, deleterious effects can impair an unshielded fiber optics cable that is combined with a high voltage power line. If, over a long period of time, the electric field produces sufficient deterioration even in just one location of a long length of fiber, the transmission of information of the entire length of the fiber will be disrupted. This could result from the high electric field that emanates from the power line, which stresses the dielectric material of the optics cable. Electrical treeing in the fibers is one degradation mechanism. Electrical treeing refers to the formation of branching structures in a dielectric due to high electric stress, and is similar to Lichtenberg figures. Electrical trees occur in the dielectric (e.g. cross-linked polyethylene) of underground transmission and distribution cables, and are related to electrochemical trees and water trees.

For a dc power line, the fiber optics cable is polarized by the electric field, and the electric stress internal to the cable can cause deterioration of its optical properties. For an ac power line, the fiber optics cable is alternately polarized in one direction and then the opposite direction as the electric field alternates. For a 60 Hz power line this change in polarization takes place 120 times per second which causes a dielectric power loss in the optics cable as well as stressing the fiber. A patent has been issued for an invention which pertains to method and apparatus for protecting the fiber optics cable from the high electric field of power lines [13]. We feel that this protection will enable fiber optics cables to be compatibly carried on power lines in the future as part of the IS.

The impact of IS on utilities can be significant in providing a new role for power/information brokers, new markets, and advanced simulation techniques needed in the control systems of the future. Updated information is essential on business opportunities and risks to help electric utilities in understanding the national information infrastructure and potential telecommunications strategies before making any major decisions. A window of both necessity and opportunity faces utilities. The necessity is to handle, process, and transmit information to survive in the present competitive milieu of electric power delivery. The opportunity is to find new sources of revenue in the new arena as well as use the new arena to advantage in the electric power field by wisely managing the supply and transmission of electricity to meet fast and widely changing demands. One important utility function that IS will help to achieve is that of Distribution Automation as discussed in the next Section.

**Distribution Automation**

Automation of distribution feeder circuits, residential loads, and commercial customer loads should decrease energy costs, allow for faster customer payments, improve power reliability and quality, provide the potential for variable-priced energy provisions and sales, and reduce utility operational costs. It will facilitate monitoring of energy use for energy management systems. Another advantage will be the ability for automatic meter reading for electricity, gas, and water. The challenge will be to automate distribution in such a way that the entire range of utilities from small to large, with differing technologies will all benefit from this innovation.

To accomplish these goals, EPRI created a Distribution Automation Pilot Project (DAPP). This includes the automation of 2 distribution substations and feeders, the automation of 20 commercial customers, and the automation of 200 commercial sites. In addition to demonstrating the benefits of DAPP, this will test our Utility Communications Architecture (UCA). UCA creates a standard, non-proprietary communications architecture whose purpose is to:

- Facilitate interoperability between different computer systems.
- Reduce product costs through standardization.
- Enable and improve compatibility of different hardware and software systems.
- Allow utility personnel to access information across the utility spectrum.
- Provide for the exchange of information between hardware systems within a utility and between two or more utilities.

Additionally, the automated distribution management functions will analyze power flow; determine connectivity; detect, locate, and isolate faults; restore service; control voltage/vars; and reconfigure feeders. In addition to automatic meter reading, other customer site functions will include load control, tamper detection, outage detection and restoration, connect/disconnect, notification of status of outage restoration (i.e. should commercial customer send employees home; should residential customer make other plans; etc.), and customer notification after power restoration. Not only could billing information be provided daily, it could be itemized by appliance or larger segments. Of course it would also permit electronic payment of bills. Real-time pricing based on actual cost of generation would allow the customer to have scheduled usage of appliances and industrial equipment. In dense urban areas, distribution automation may be facilitated by fiber optic plus coaxial cable. In less dense rural areas, this can be done by less expensive radio communication.

**Conclusion**

If novel power systems are to be incorporated in electric utilities, they must either fill a new niche, or compete both technically and economically with already well developed systems. Innovation is difficult to achieve for any industry that has become highly technical and capital intensive over a century of development, as is the electric power industry. We should be careful to avoid either of two extremes. We shouldn't reject pursuing new technologies just because they seem alien and unfamiliar to us. Neither should we blindly accept a new technology simply because it has received a lot of media exposure and is the latest fad. As we look at power systems of the future, new technologies often appear more promising than they turn out to be, precisely because they are remote. Their warts are not perceptable at a distance. Their drawbacks and flaws only become evident as we see them more closely.  Occasionally, the remoteness of a technology leads to unduly pessimistic conclusions about its future.  Even after a technology has been demonstrated, leading scientists may have doubts about its practicality because of the necessity for new developments which are needed, but cannot be foreseen.  So with any evaluation of future delivery technologies, new developments may well alter presently sound conclusions.

Although dispersed generation using renewable energy sources may not impact within the next two decades, it is not only inevitable, but regulation may bring about it's commercial emergence much sooner than most expect.  There was much truth in what Malthus said about populations and needs tending to increase geometrically whereas resources tend to increase only arithmetically, leading to crises.   Malthus overlooked two factors that have thus far vitiated his conclusion:  societal and technological change.  Of these two, technological innovation has played the more important role.  Given the present rate of population growth and the per capita increased demand for energy as the less developed countries improve their standard of living, it is fairly clear that even if we had easy access to all the oil, gas, and coal in the earth, these resources would all be consumed in about 100 to 300 years from now.  We will certainly encounter crises well before this if alternate forms of energy are not accepted into our societal infrastructure and substantially incorporated into its power system.

There is truth in economy of scale, that operation below a critical size is wasteful. Power delivery and power production both have economies of scale.  Well before limits are reached it is roughly like a 2/3 power law for the overall system, as is a surface to volume ratio.  This is because costs related to permanent materials scale like a surface, and costs of consumable energy materials scale like a volume.  The factor that is neglected in this simplistic view is the increased failure mode probability relative to smaller redundant systems if the scale gets too large.  Dispersed generation may provide a new framework for the power delivery system that has the potential of increasing its reliability.  This will work well in coordination with distribution automation, which will allow better control of both the  distribution system and its loads.

Financial and environmental pressures have forced more intensive utilization of available power delivery.  Scoping studies concluded that FACTS can produce significant savings for scenarios in which utilities benefit from improved control of power flow (avoiding loop flow) and in situations which are stability limited.  It is self-evident that the Custom Power aspect of FACTS is a vital asset in producing and delivering quality power.   Avoidance of building new lines is a major benefit, when the existing system reaches its power delivery limit.  Even though  FACTS  will increase the total power use-capacity, it will decrease neither the absolute amount of power losses nor their relative percentage. These will increase both in the lines and in the ancillary equipment.

Hyperconductivity and eventually high temperature superconductivity may be ways to reduce power losses and increase power density. Our existing grid system is a valuable resource and FACTS can help us make the most of that resource. New lines are expensive and require time and resources for permits and construction, and the issuance of permits is by no means certain.

In terms of increasing power delivery capacity, there is an average limit of about 20% as to what FACTS can do, so we must also look to new technologies. It may be a while before the problems of brittleness and critical current density will be sufficiently solved for high temperature superconductors (HTSC) that it can have much of a near-term impact. It is difficult to say whether in the near-term HTSC will be able to fill a niche in the retrofitting of 3-in-1 pipe type cables, where the need for increased power carrying capacity overrides the need for reduced losses.

Hyperconductivity using beryllium at 77 K looks very promising to both increase power-carrying capacity and reduce losses. We are aware of the problems that are a deterrent to use of Be, and we should properly take a cautious approach in evaluating its potentiality. Would $SF_6$ ever have been used if it had been known in advance that $S_2F_{10}$, an arcing by-product, is one of the most poisonous gases known? By serendipity, this didn't turn out to be a problem. We should at least do some of the preliminary research to ascertain if Be's technical potential can be achieved in power lines, and if its cost can be significantly reduced in case there is a greatly increased demand for it.

The information superhighway presents an unparalleled opportunity for electric utilities in the emerging deregulated (or perhaps more correctly re-regulated) environment. Sophisticated customers may have computer input and control to make frequent changes in which utility will be delivering power to them. One possible implication is that as a function of market-driven usage some lines will need to operate at limit capacity, and some that are now well-utilized may become under-utilized. In order to survive in this new milieu, utilities will have to become highly competitive in the price of delivered power and to anticipate these kinds of changes. With respect to failure rates, utilities will need to ascertain which of their lines are the worst performers. Upgrading of lines should take failure rates into consideration as well as anticipated increased usage.

It appears that in changing the very nature of the U.S. electric utility industry, the regulators may not have considered all the implications -- from destabilizing the power grid to undermining millions of innocent shareholders who have traditionally relied on utility stocks as an instrument of reliable and stable investment. A nationwide comprehensive analysis is needed to ascertain the full implications of this new policy. Among the issues that need greater clarification are:

1. Full implication of the economic consequences with respect to the present policy, and a more moderate re-regulation policy.

2. National security implications of this new policy.

3. Grid stability implications.

4. New generation risks from large load transients (e.g. turbogenerator shaft vulnerability).

5. Environmental impact.

Because of the unavailability of new rights-of-way for overhead lines, it is clear that an increasing amount of power delivery will have to be underground. As the available corridors become saturated and power dissipation increases as fast or faster than the increase in capacity, more attention will need to be given to the thermal conductivity of the backfill. A slack wax was developed which can stabilize the thermal conductivity of the soil. Slack wax is an inexpensive by-product of oil refining that is stable in the ground, and can be added to backfill in emulsified form, or by heating.

Let us hope that the present underground vault explosions are not a harbinger of worse to come. Programs are focused on determining the cause, and preventing the explosions. In the worst case scenario, we may have to develop insulations that are hydrocarbon-free. A major R & D effort would be required to commercialize cables using such new dielectrics in the next 20 years.

Global competition requires that R & D results be moved into the marketplace for utilization by electric power companies with as much care and speed as possible.